\documentclass[twocolumn]{webofc}

\usepackage[varg]{txfonts}   
\usepackage{hyperref}
\usepackage{url}
\usepackage{bbold}

\hypersetup{colorlinks=true,citecolor=blue,urlcolor=blue,linkcolor=blue}

\begin{document}

\title{Impact of shell model interactions on WIMP-nucleus scattering observables for silicon and germanium targets}

\author{\firstname{Raghda} \lastname{Abdel Khaleq}\inst{1,2,3}\fnsep\thanks{\email{raghda.abdelkhaleq@anu.edu.au}} \and
        \firstname{Madeleine} \lastname{J. Zurowski}\inst{4}\fnsep\thanks{\email{madeleine.zurowski@utoronto.ca}}   
}

\institute{Department of Fundamental and Theoretical Physics, Research School of Physics, Australian National University, ACT, 2601, Australia 
\and
           Department of Nuclear Physics and Accelerator Applications, Research School of Physics, Australian National University, ACT, 2601, Australia
\and 
            ARC Centre of Excellence for Dark Matter Particle Physics, Australia
\and
           Department of Physics, University of Toronto, Toronto, ON M5S 1A7, Canada
          }

\abstract{
The nature of dark matter (DM) remains one of the biggest mysteries in physics today. Dark matter direct detection experiments look for nuclear recoil signals from DM-nucleus elastic scattering, which can be used to characterise DM. Nuclear modelling of the target nucleus may impact the predicted DM-nucleus scattering rates, and affect interpretation of experimental signals. In this work, we investigate the impact of nuclear shell model interactions on DM nuclear responses for silicon and germanium targets using a SuperCDMS-like experimental parameters. Nuclear uncertainties resulting from shell model interaction choice in the nuclear form factors are roughly retained at the scattering rate and exclusion limit levels for certain nuclear responses. 
}

\maketitle

\section{Introduction}
\label{intro}

Understanding the nature of dark matter (DM) has been a long-standing goal in physics, spanning many decades. In response, a plethora of DM particle candidates have been proposed, one class being the Weakly Interacting Massive Particle (WIMP). WIMPs are expected to have a weak-scale mass in the range of $\sim$ GeV/$c^2$$-$TeV/$c^2$, and are non-relativistic (NR) with velocities $\sim10^{-3}c$.
Dark matter direct detection experiments aim to study the nature of DM through signals obtained from a WIMP elastically scattering off a target nucleus in a detector, causing this nucleus to recoil. The DM-nucleus scattering rate consists of several components, each which must be modelled accurately -- uncertainties present in any of these will impact the interpretation and analysis of experimental signals. 
These components include: the DM halo velocity distribution; the high energy physics content employed; response functions describing the direct detection experiment of interest; and the target nuclear structure information. \\

Thus far, direct detection experiments have employed a range of nuclear targets in order to better characterise experimental signals. The WIMP-nucleus interaction may be sensitive to different aspects of the nuclear structure, which is unique to each of the experimental targets. 
As such, accurate characterisation of the nuclear structure information and its impact on scattering rate predictions are important. The standard characterisation of the WIMP nuclear response involves considering both spin-independent (SI) and spin-dependent (SD) responses. The SI response receives a coherent enhancement proportional to the atomic mass in the form of $A^2$, and is hence typically expected to be the largest response with the strongest limits. By accounting for the motion of nucleons within the nucleus, additional momentum-suppressed terms are included in the WIMP nuclear responses. The work of  Fitzpatrick \textit{et al}. \cite{Fitzpatrick:2012ix,Anand:2013yka} employed a non-relativistic effective field theory (NREFT) approach to consider these additional nuclear responses, which depend on orbital angular momentum $L$ (LD) as well as both spin $S$ and $L$ (LSD). In this work, we employ this NREFT framework to investigate the impact of nuclear structure on silicon and germanium targets for SuperCDMS-like experimental parameters. This is performed using large-scale nuclear shell model calculations.

\section{Scattering Rate}

The DM-nucleus elastic scattering rate, expressed in cpd (counts per day) per kilogram per keV, is given by

\begin{equation}\label{standard rate expression}
    \frac{{\rm d} R_T}{{\rm d}E_R} = \frac{N_T \rho_\chi}{m_\chi} \int_{v>v_{\text{min}}} v f(\vec{v}) \ \frac{{\rm d}\sigma_T}{{\rm d}E_R} (\vec{v}) \ {\rm d}^3v,
\end{equation}

\noindent where $m_\chi$ is the DM mass, $E_R$ is the nuclear recoil energy, and $N_T$ is the number of target nuclei of type $T$ per detector mass. ${\rm d}\sigma_T/{\rm d}E_R$ is the differential cross section, which holds information about the nuclear physics of the target nuclei, as well as the particle physics content through high energy coefficients. Here, $f(\vec{v})$ is the WIMP halo velocity distribution in the Earth reference frame, $\rho_\chi$ is the local DM density, and a sum $\sum_T N_T \ {\rm d}\sigma_{T}/{\rm d}E_R$ is performed for targets with more than one isotope/nucleus. The minimum DM velocity required to produce a nuclear recoil $E_R$ is given by 

\begin{equation}
    v_{\text{min}} (E_R) = \frac{q}{2\mu_T} = \frac{1}{\mu_T} \sqrt{\frac{m_T E_R}{2}},
\end{equation}

\noindent where $\mu_T= m_T m_\chi/(m_T+m_\chi)$ is the DM-nucleus reduced mass and $m_T$ is the nuclear target mass. The WIMP-nucleon momentum transfer is given by $\vec{q}= \vec{p}\hspace{0.5mm}'-\vec{p}= \vec{k}-\vec{k}'$, with $\vec{p}\hspace{0.5mm}'$ ($\vec{p}$) being the outgoing (incoming) $\chi$ momentum and $\vec{k}'$ ($\vec{k}$) the outgoing (incoming) $N$ momentum. \\

In this work the DM halo velocity distribution is taken to be the standard halo model (SHM), which is modelled as an isotropic halo distribution with the form \cite{Evans:2018bqy,OHare:2019qxc}

\begin{equation}\label{eq:shm}
  f(\vec{u}) = \frac{1}{(2\pi
    \sigma_v^2)^{3/2}N_\mathrm{R,esc}} \, \exp \left( - \frac{|\vec{u}|^2}{2\sigma_v^2}\right) \Theta (v_{\rm esc} - |\vec{u}|)\,, 
\end{equation}

\noindent where

\begin{equation}
    N_\mathrm{esc} = {\rm erf} \left( \frac{v_{\rm esc}}{\sqrt{2}\sigma_v}\right) -
  \sqrt{\frac{2}{\pi}} \frac{v_{\rm esc}}{\sigma_v} \exp \left(
  -\frac{v_{\rm esc}^2}{2\sigma_v^2} \right)\,.
\end{equation}

The Earth reference frame velocity distribution $f(\vec{v})$ can be obtained through a Galilean transformation $f(\vec{v})=f(\vec{u}+\vec{v}_E)$, where $\vec{v}_E$ is the Earth velocity with respect to the galactic frame. Here, $N_\mathrm{R,esc}$ is the normalisation factor, and $\sigma_v$ is the DM isotropic velocity dispersion with $v_0 = \sqrt{2} \sigma_v$. The model parameters employed here are $\rho_\chi =0.3$ GeV/$c^2$ cm$^{-3}$, $v_0 = 220$ km s$^{-1}$, and $v_{\rm{esc}} = 544$ km s$^{-1}$ \cite{Evans:2018bqy}. \\

The differential cross section can be written as

\begin{equation}\label{anand cross section}
     \frac{{\rm d}\sigma_T}{{\rm d}E_R} = \frac{m_T}{2\pi v^2} \sum_{i,j} \sum_{N,N'=p,n} \mathfrak{c}_i^{(N)} \mathfrak{c}_j^{(N')} F_{i,j}^{(N,N')} (v^2,q^2),
\end{equation}

\noindent where $\mathfrak{c}_i^{(N)}$ are the high energy coefficients which hold information about the particle physics content of the model employed. The nuclear form factors $F_{i,j}^{(N,N')}$ can be obtained from the NREFT Hamiltonian for this scattering process, where they hold information about the specifics of the nuclear structure. The nuclear uncertainty can be studied through these form factors. Each $F_{i,j}^{(N,N')}$ consists of a combination of the form factors $F^{(N,N')}_{X,Y}$, where \cite{Fitzpatrick:2012ix,Anand:2013yka}

\begin{equation}\label{FF eqn}
    F^{(N,N')}_{X,Y} (q^2)  \equiv \frac{4\pi}{2J_i+1}  \sum_{J=0}^{2J_i} \langle J_i || X_J^{(N)} || J_i \rangle \langle J_i || Y^{(N')}_J || J_i \rangle.
\end{equation}

Here, $N, N'=\{p, n\}$ denote the proton and neutron components, where $X_J^{(p)}= \frac{1+\tau_3}{2} \ X_J$ and $X_J^{(n)}= \frac{1-\tau_3}{2} \ X_J$, with $\tau_3$ begin the nucleon isospin operator. For the non-interference responses (with $X=Y$) we have $F^{(p,n)}_{X,Y} (q^2)= F^{(n,p)}_{X,Y} (q^2)$, and we define $F^{(N,N')}_X (q^2)  \equiv  F^{(N,N')}_{X,X} (q^2)$. The explicit definition of each $F_{i,j}^{(N,N')}$ in terms of $F^{(N,N')}_{X,Y}$ can be found in Refs.~\cite{Fitzpatrick:2012ix,Anand:2013yka}. $X,Y$ denote one of six nuclear operators $M_{JM}$, $\Sigma''_{JM}$, $\Sigma'_{JM}$, $\Delta_{JM}$, $\Phi''_{JM}$ and $\tilde{\Phi}'_{JM}$. Each of these corresponds to a different aspect of the WIMP-nucleus scattering process, with $M_{JM}$ describing the standard SI response, and $\Sigma''_{JM}, \ \Sigma'_{JM}$ the Longitudinal and Transverse SD responses, respectively. Additional momentum-suppressed responses are now included -- namely an orbital angular momentum dependent (LD) response $\Delta_{JM}$; a spin-orbit response $\Phi''_{JM}$; and a complex tensor response $\tilde{\Phi}'_{JM}$. Both $\Phi''_{JM}$ and $\tilde{\Phi}'_{JM}$ depend on angular momentum and spin, and are hence referred to as LSD responses. The explicit definitions of the nuclear operators $X$ can be found in Refs.~\cite{Fitzpatrick:2012ix,Anand:2013yka}.  \\

These nuclear channels correspond to a combination of non-relativistic (NR) operators $\mathcal{O}^{\rm NR}_i$, which enter into the WIMP-nucleus interaction Lagrangian as

\begin{equation}\label{Fitz:LagrangianGeneral}
    \mathcal{L}_{\text{int}}^{\text{NR}} = \sum_{N=n, p} \sum_{i} \mathfrak{c}_i^{(N)} \mathcal{O}^{\text{NR}}_i \ \chi^+ \chi^- N^+ N^-,
\end{equation}

\noindent where $\chi$ represents the dark matter field and $N$ a nucleon field. In the NR regime the only terms kept are those which depend on $\vec{q}$ up to second order. These NR operators have the form 

\begin{equation}
\label{fitz_nreft}
\begin{aligned}
\mathcal{O}^{\rm NR}_1 &= \mathbb{1} \ ,
&
& ,
\\
\mathcal{O}^{\rm NR}_3 &= i \, \vec{S}_N \cdot \left(\vec{q} \times \vec{v}^\perp \right) \ ,
&
\mathcal{O}^{\rm NR}_4 &= \vec{S}_\chi \cdot \vec{S}_N \ ,
\\
\mathcal{O}^{\rm NR}_5 &= i \, \vec{S}_\chi \cdot \left(\vec{q} \times \vec{v}^\perp \right) \ ,
&
\mathcal{O}^{\rm NR}_6 &= \left(\vec{S}_\chi \cdot \vec{q}\right) \left(\vec{S}_N \cdot \vec{q}\right) \ ,
\\
\mathcal{O}^{\rm NR}_7 &= \vec{S}_N \cdot \vec{v}^\perp \ ,
&
\mathcal{O}^{\rm NR}_8 &= \vec{S}_\chi \cdot \vec{v}^\perp \ ,
\\
\mathcal{O}^{\rm NR}_9 &= i \, \vec{S}_\chi \cdot \left(\vec{S}_N \times \vec{q} \right) \ ,
&
\mathcal{O}^{\rm NR}_{10} &= i \, \vec{S}_N \cdot \vec{q} \ ,
\\
\mathcal{O}^{\rm NR}_{11} &= i \, \vec{S}_\chi \cdot \vec{q} \ ,
&
\mathcal{O}^{\rm NR}_{12} &= \vec{v}^\perp \cdot \left(\vec{S}_\chi \times \vec{S}_N \right) \ ,
\end{aligned}
\end{equation}

\noindent where the terms included are those which are at most quadratic in either the spin $\vec{S}$ or velocity $\vec{v}$. Here, the relative velocity is defined as $\vec{v}^{\perp} \equiv \vec{v} +\vec{q}/(2 \mu_N)$, where $\vec{v}_T^\perp \equiv \vec{v}_T+\vec{q}/(2\mu_T)$. Table~\ref{table of NR and q operator nuclear} presents a summary of the dependence of the NR operators $\mathcal{O}^{\rm NR}_i$ on the nuclear channels $X$, where a factor of ${v_T^\perp}^2$ or $q^2$ in the cell indicates a dependence of the form ${v_T^\perp}^2 X$ or $q^2 X$.

\begin{table}[htbp] 
\centering
\caption{Summary of the dependence of the $O_{i}^{\rm NR}$ operators on the nuclear responses $M$, $\Sigma''$, $\Sigma'$, $\Delta$, $\Phi''$, $\tilde{\Phi}'$. The suppression factor (${v_T^\perp}^2$ and/or $q^2$) is also presented, with $1$ representing no suppression, and an empty cell indicating no dependence on the nuclear response. 
\label{table of NR and q operator nuclear}}
\resizebox{0.8\linewidth}{!}{%
\begin{tabular}{|c||c|c|c|c|c|c|}
\hline 
          \textbf{} & $M$ & $\Sigma''$ & $\Sigma'$ & $\Delta$ & $\Phi''$ & $\tilde{\Phi}'$  \\
\hline 
\hline
               \textbf{$\mathcal{O}_1^{\rm NR}$} & 1 &  &  &  &  &   \\
\hline
             \textbf{$\mathcal{O}_3^{\rm NR}$} &  &  & ${v^\perp_T}^2  q^2$ &  & $q^4$ &   \\
\hline
             \textbf{$\mathcal{O}_4^{\rm NR}$} &  & 1 & 1  &  &  &   \\
\hline
             \textbf{$\mathcal{O}_5^{\rm NR}$} & ${v^\perp_T}^2 q^2$ &  &  & $q^4$ &  &   \\
\hline
             \textbf{$\mathcal{O}_6^{\rm NR}$} &  & $q^4$ &  &  &  &   \\
\hline
             \textbf{$\mathcal{O}_7^{\rm NR}$} &  &  & ${v^\perp_T}^2$ &  &  &   \\
\hline
             \textbf{$\mathcal{O}_8^{\rm NR}$} & ${v^\perp_T}^2$ &  &  & $q^2$ &  &   \\
\hline
             \textbf{$\mathcal{O}_9^{\rm NR}$} &  &  & $q^2$ &  &  &   \\
\hline
             \textbf{$\mathcal{O}_{10}^{\rm NR}$} &  & $q^2$ &  &  &  &   \\
\hline
             \textbf{$\mathcal{O}_{11}^{\rm NR}$} & $q^2$ &  &  &  &  &   \\
\hline
             \textbf{$\mathcal{O}_{12}^{\rm NR}$} &  & ${v^\perp_T}^2$ & ${v^\perp_T}^2$ &  & $q^2$ & $q^2$   \\
\hline
\end{tabular}
}
\end{table}

\section{Shell Model Calculations}
Large-scale shell model calculations were performed using NuShellX \cite{Brown:2014bhl}, which employs a spherical shell model basis. For each valence (model) space considered, the shell model interaction is varied in the program, hence resulting in different form factor values. Through these different shell model interactions we quantify the nuclear uncertainties present in the nuclear form factors and scattering observables. \\

Calculations for the silicon isotopes $^{28,29,30}$Si were preformed in the full $sd$ model space with single particle levels $1d_{5/2}, \ 2s_{1/2},\ 1d_{3/2}$ for both protons and neutrons. Two shell model interactions were employed, USD \cite{Brown:1988vm, Wildenthal:1984mf} and USDB \cite{Brown:2006gx}, where the latter is an upgraded iteration of the former. Theoretical shell model values for the $^{29}$Si energy levels and electromagnetic moments and transitions were compared to the experimental counterparts, and can be found in Ref.~\cite{AbdelKhaleq:2023ipt}. \\

Calculations for the germanium isotopes $^{70,72,73,74,76}$Ge were performed in the full $f_5pg_9$ valence space, with single-particle levels $2p_{3/2}$,  $1f_{5/2}$, $2p_{1/2}$, and $1g_{9/2}$. The form factors were evaluated for two shell model interactions, JUN45 \cite{Honma:2009zz} and jj44b \cite{Mukhopadhyay:2017tca}. These were compared against the results of Fitzpatrick \textit{et al.} \cite{Fitzpatrick:2012ix,Anand:2013yka}, where the  GCN2850 interaction \cite{Menendez:2008jp} was used, with a valence space truncation where the occupation number of the $1g_{9/2}$ level is limited to no more than two nucleons above the minimum occupation for all isotopes. The experimental energy levels and electromagnetic moments and transitions are compared against the theoretical jj44b and JUN45 values in Refs.~\cite{AbdelKhaleq:2023ipt,AbdelKhaleq:2024hir} for all germanium isotopes. \\

In Ref.~\cite{AbdelKhaleq:2023ipt} these shell model calculations were employed to calculate the nuclear form factors $F^{(N,N')}_{X} (q^2)$ for each shell model interaction, which were compared against one another to quantify the nuclear uncertainty present. This comparison was facilitated through an integrated form factor (IFF) value for each nuclear channel $X$, which acts as a gauge for the strength of each channel as well as the magnitude of the sensitivity of the form factors to shell model interaction choice. The extensive nuclear uncertainty discussion for silicon and germanium can be found in the aforementioned paper.

\section{Experimental Functions and Rates}
To evaluate rates and exclusion curves for a DM direct detection experiment, the interaction rate in Eq.~(\ref{standard rate expression}) must be considered alongside functions which describe responses specific to that experiment, such as efficiency, acceptance, and energy resolution. In this work we model the signal assuming detector response values reported by SuperCDMS \cite{Aralis_2020}.
The SuperCDMS SNOLAB experiment employs silicon (Si) and germanium (Ge) target nuclei, and is located in Sudbury, Canada. Two types of cryogenic detectors are used, namely High Voltage (HV) and Interleaved Z-sensitive Ionisation and Phonon (iZIP) detectors \cite{supercdmscollaboration2023}. The calculations here are restricted to the HV detectors. The detector's energy scale, in terms of the recoil energy $E_R$, is given by 

\begin{equation}
    E_{\rm ph} = E_R\left( 1 + \frac{y(E_R)}{\varepsilon_{\rm eh}}eV \right),
\end{equation}

\noindent where $E_{\rm ph}$ is the observed phonon energy, $y(E_R)$ is the ionisation yield, $\varepsilon_{\rm eh}$ is the energy required to produce a single electron-hole pair, and $V$ is the voltage applied to the detector. The observed energy is written as

\begin{equation}
    E_{\rm obs} = E_R\left( \frac{\varepsilon_{\rm eh} + y(E_R)eV}{\varepsilon_{\rm eh}+ eV} \right),
\end{equation}

\noindent where the observation rate then has the form

\begin{equation}
    \frac{{\rm d}R}{{\rm d}E}=\frac{\epsilon(E)}{(2\pi)^{1/2}}\int_{0}^{\infty}\frac{1}{\Delta E_{\rm obs}}\frac{{\rm d}R}{{\rm d}E_{\rm obs}}\exp\left[\frac{-(E-E_{\rm obs})^2}{2(\Delta E_{\rm obs})^2}\right]{\rm d}E_{\rm obs}.
\end{equation}

Here, $\epsilon(E)$ is the efficiency factor as a function of energy, and a Gaussian resolution of the form $\Delta E_{\rm obs} = \sqrt{DE_{\rm obs}+(CE_{\rm obs})^2 + \sigma_E^2}$ is employed. This resolution is approximated by that of CDMSlite \cite{Aralis_2020}, where for both germanium and silicon targets we take  $C = 5\times 10^{-3}$ and $D = 0.7 \text{ eV}$. Additionally, the values $\sigma_E (\rm{Si}) = 5 \text{ eV}$ and $\sigma_E (\rm{Ge}) = 10 \text{ eV}$ are employed, with an 85\% flat efficiency for both nuclear targets. The ionisation yield expressions are given by \cite{Barker_2013, si_yield} 

\begin{equation}
    y(E_R)_{\rm{Ge}} = \frac{k~g(\alpha)}{1+k~g(\alpha)}, \hspace{5mm} y(E_R)_{\rm{Si}} = y_{10{\rm keV}}\left(\frac{E_R}{10{\rm keV}}\right)^B,
\end{equation}

\noindent where the germanium parameters are 

\begin{equation}
\begin{split}
    & k = 0.133 Z^{2/3}A^{-1/2}, \hspace{5mm} \alpha = 11.5E_R Z^{-7/3}, \\
    & g(\alpha) = 3\alpha^{0.15} + 0.7 \alpha^{0.6} + \alpha,
\end{split} 
\end{equation}

\noindent and the silicon counterparts are

\begin{equation}
\begin{split}
    y_{10{\rm keV}} = 0.302, \hspace{5mm} B = 0.261.
\end{split}    
\end{equation}

The silicon and germanium total detector masses are 2.4 kg and 11.12 kg, respectively. The background for both nuclei is taken to be flat, with a value of 0.1 cpd/kg/keV for silicon and 0.003 cpd/kg/keV for germanium. Pseudo-data were employed for the energy region 0.1-10 keV assuming a total run time of 5 years, where the Optimal Interval method \cite{Yellin:2002xd} was used for the sensitivity projections. A python code was developed to calculate the rates for a range of target nuclei and NR operators \cite{ZurowskiGitHubSuperCDMS}.

\section{Results}

The scattering rate and exclusion limit results are presented for germanium in Sec.~\ref{sec:GeResults} and silicon in Sec.~\ref{sec:SiResults}. The scattering rates are all plotted for $m_\chi=10$ GeV/$c^2$ and $\sigma_p=10^{-40}$ cm$^2$. Here, the high energy coefficients are set so that each NR operator is considered separately. 
The nuclear uncertainty discussion is only presented for a selection of NR operators $\mathcal{O}^{\rm NR}_i$ for brevity, however the remainder showcase similar behaviour.
The scattering observable results are compared against the IFF nuclear uncertainties presented in Ref.~\cite{AbdelKhaleq:2023ipt}, where the propagation of the nuclear uncertainty is discussed.

\subsection{Germanium}\label{sec:GeResults}

\begin{figure}[htpb]
\centering
\includegraphics[width=4cm,clip]{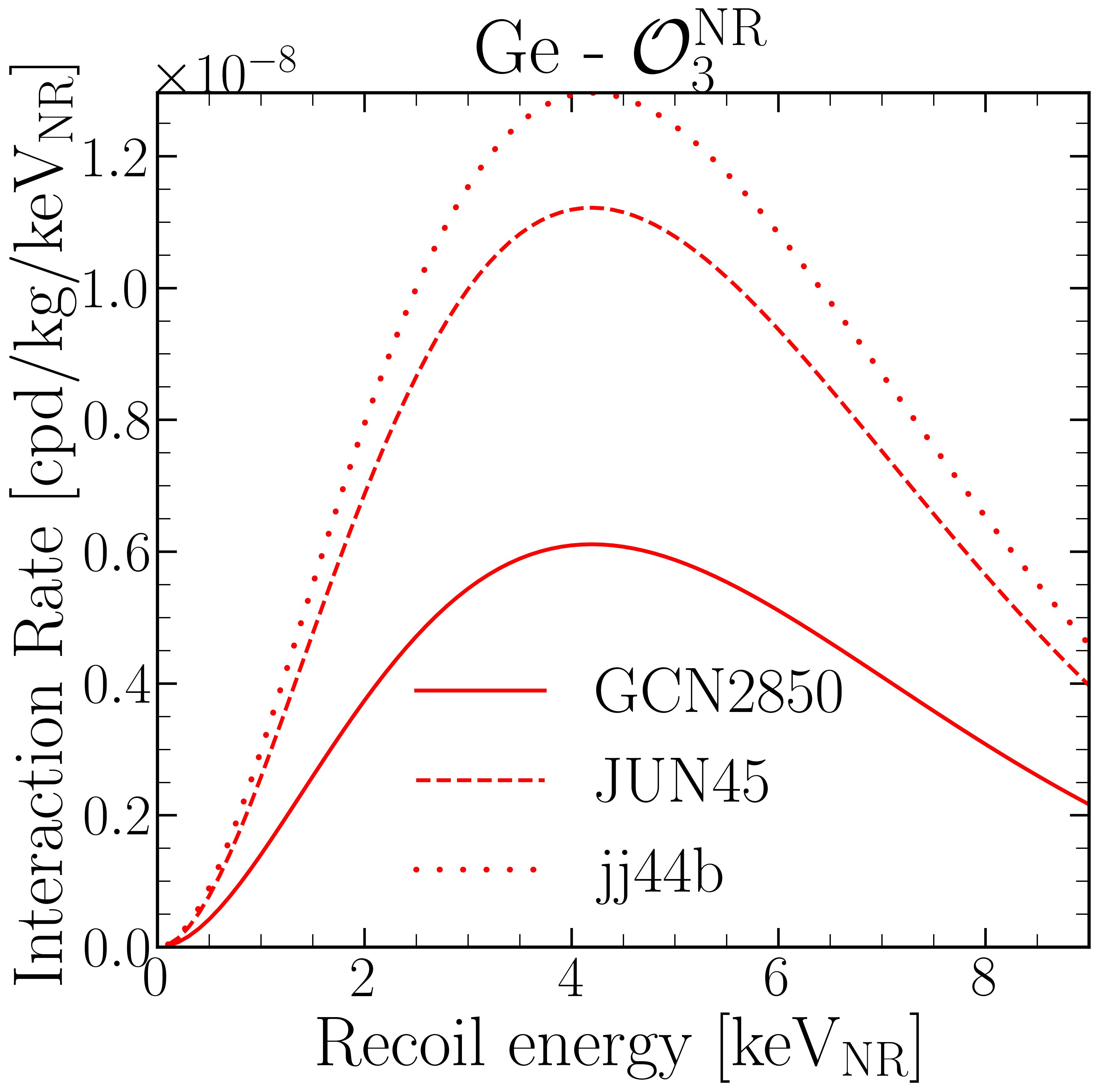}
\includegraphics[width=3.9cm,clip]{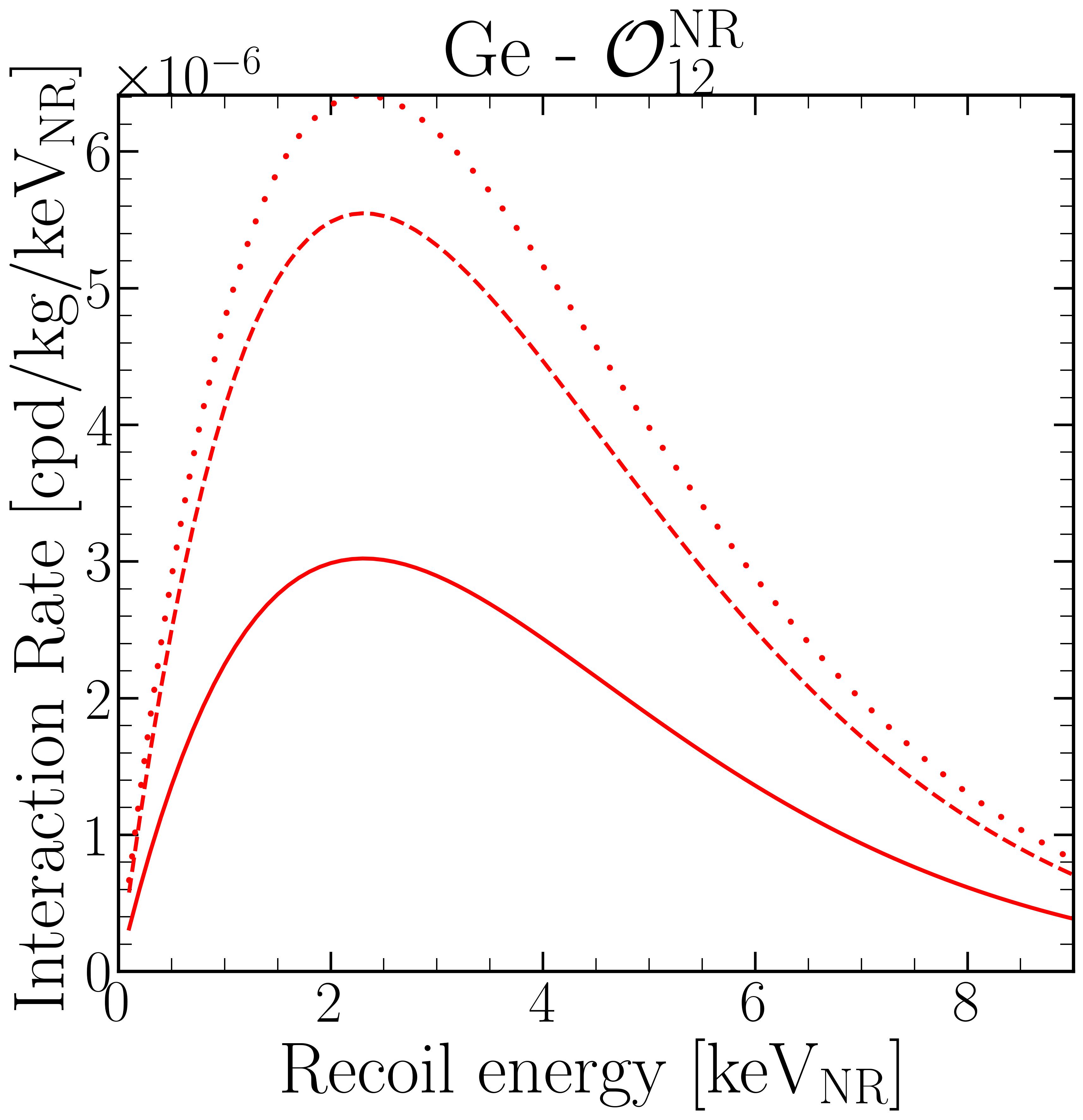}
\caption{Germanium interaction rates for the NR operators $\mathcal{O}_3^{\rm NR}$ and $\mathcal{O}_{12}^{\rm NR}$, for the shell model calculations GCN2850 (solid), JUN45 (dashed) and jj44b (dotted). Plotted using $m_\chi=10$ GeV/$c^2$ and $\sigma_p=10^{-40}$ cm$^2$.}
\label{fig: Ge NR rates}
\end{figure}

Figure~\ref{fig: Ge NR rates} presents the germanium interaction rates for the NR operators $\mathcal{O}_3^{\rm{NR}}$ and $\mathcal{O}_{12}^{\rm{NR}}$. The $\mathcal{O}_3^{\rm{NR}}$ operator consists of an SD response in addition to a spin-orbit one ($\Phi''$), whereas $\mathcal{O}_{12}^{\rm{NR}}$ consists of all SD and LSD responses. The largest germanium scattering rate differences due to nuclear shell model interaction choice exist between the GCN2850 and the jj44b calculations for the $\mathcal{O}_3^{\rm{NR}}$ and $\mathcal{O}_{12}^{\rm{NR}}$ operators. In this case, the nuclear difference has a factor of $\sim 2$, which reflects the nuclear uncertainties in the IFF values for the SD and LSD channels presented in Ref.~\cite{AbdelKhaleq:2023ipt}. As such, there is a retention of the nuclear uncertainties at the scattering rate level, particularly showcasing the importance of the LSD channel $\Phi''$ in quantifying these uncertainties.

\begin{figure}[htpb]
\centering
\includegraphics[width=4.cm,clip]{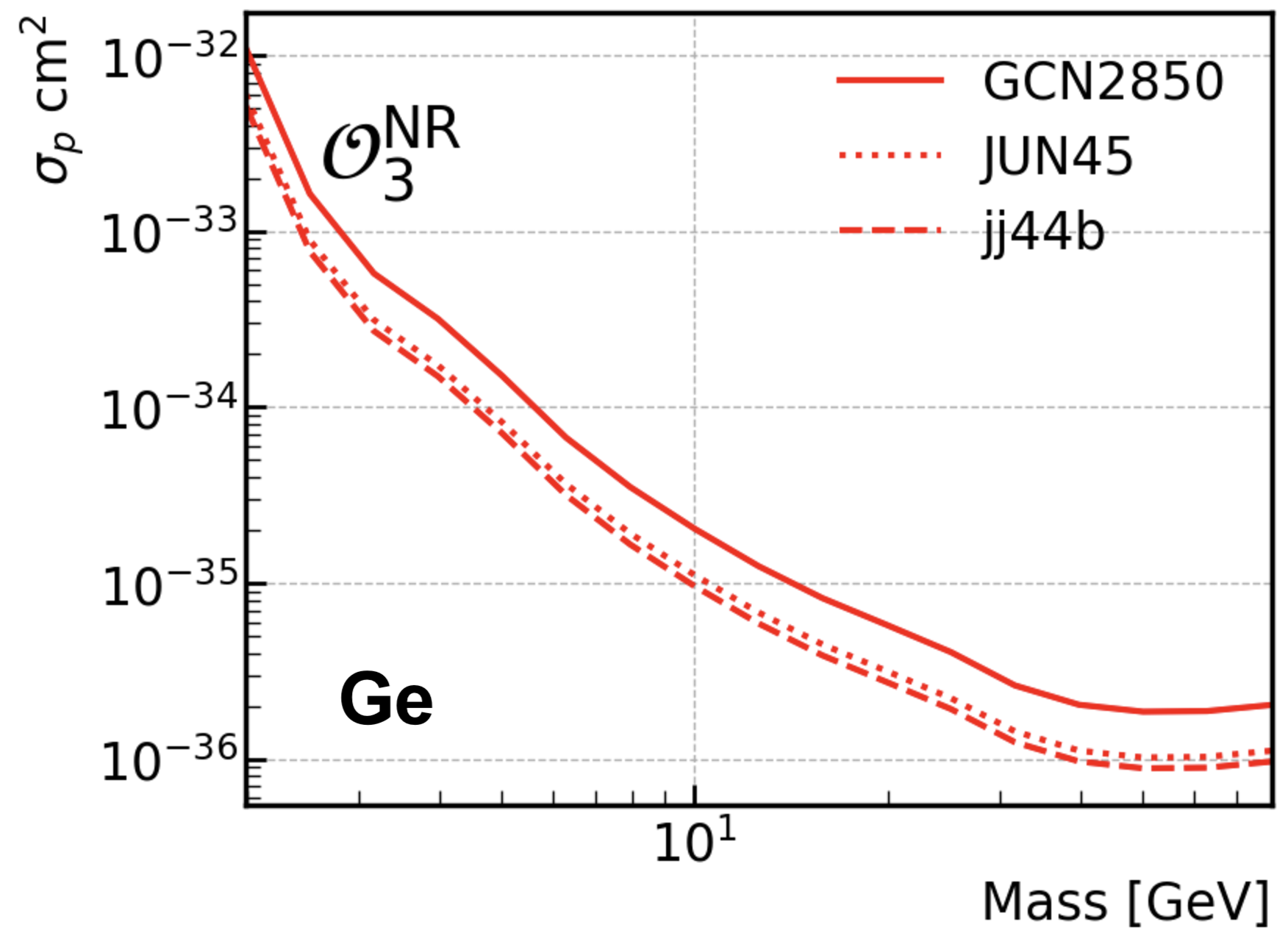}
\includegraphics[width=4.cm,clip]{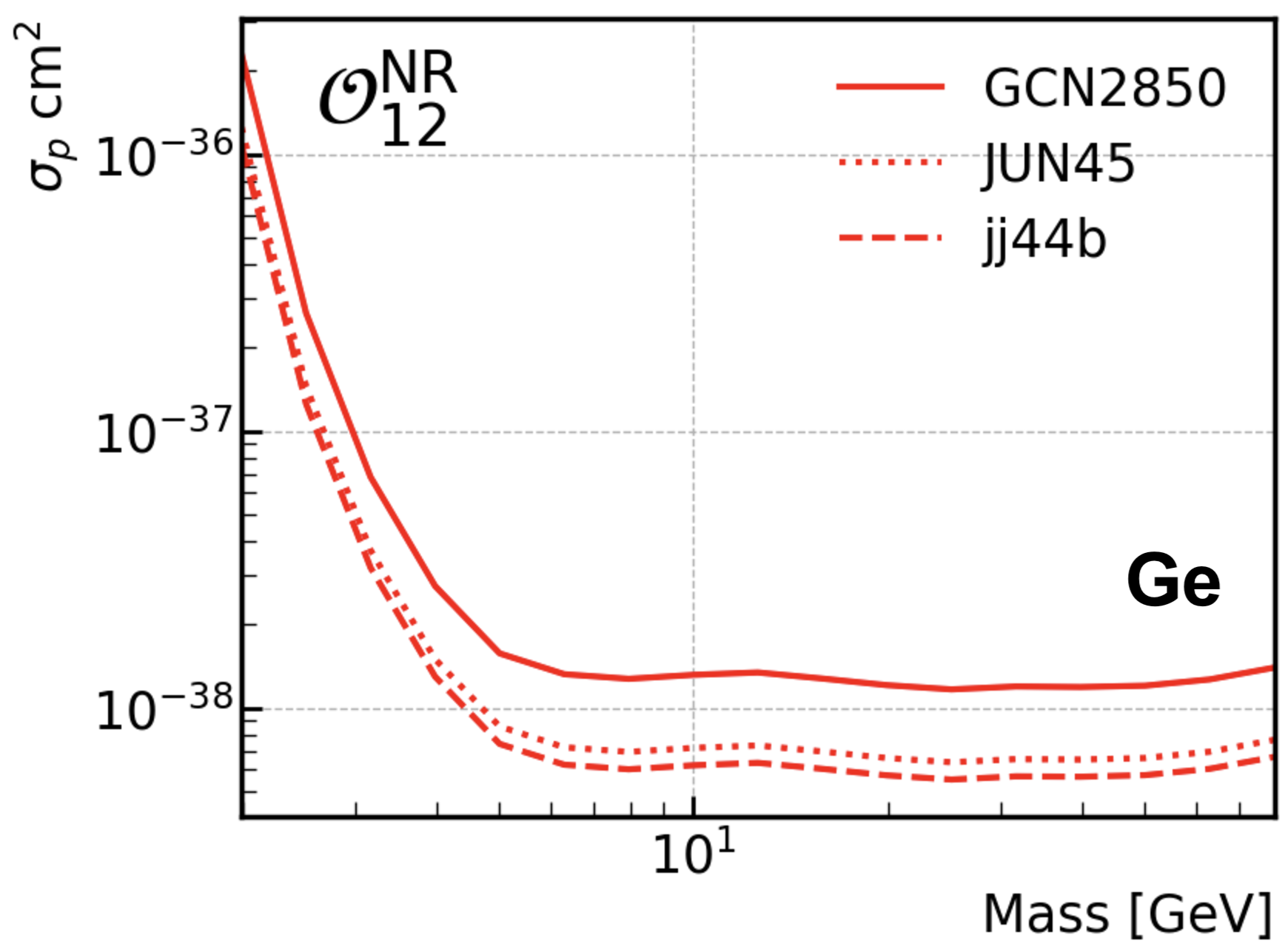}
\caption{SuperCDMS-like germanium $\mathcal{O}_3^{\rm NR}$ and $\mathcal{O}_{12}^{\rm NR}$ projections curves for three shell model calculations.}
\label{fig: ge exclusion}       
\end{figure}

These nuclear uncertainties propagate to the exclusion curves, presented in Fig.~\ref{fig: ge exclusion} for the germanium $\mathcal{O}_3^{\rm{NR}}$ and $\mathcal{O}_{12}^{\rm{NR}}$ operators. The largest differences are observed between the GCN2850 and jj44b shell model calculations, where a factor of $\sim 2$ uncertainty is present for $\mathcal{O}_3^{\rm NR}$ and a factor of $\sim 2.5$ for $\mathcal{O}_{12}^{\rm NR}$. The exclusion limits roughly retain the nuclear differences seen in the scattering rates of these NR operators. 

\subsection{Silicon}\label{sec:SiResults}

\begin{figure}[htpb]
\centering
\includegraphics[width=4cm,clip]{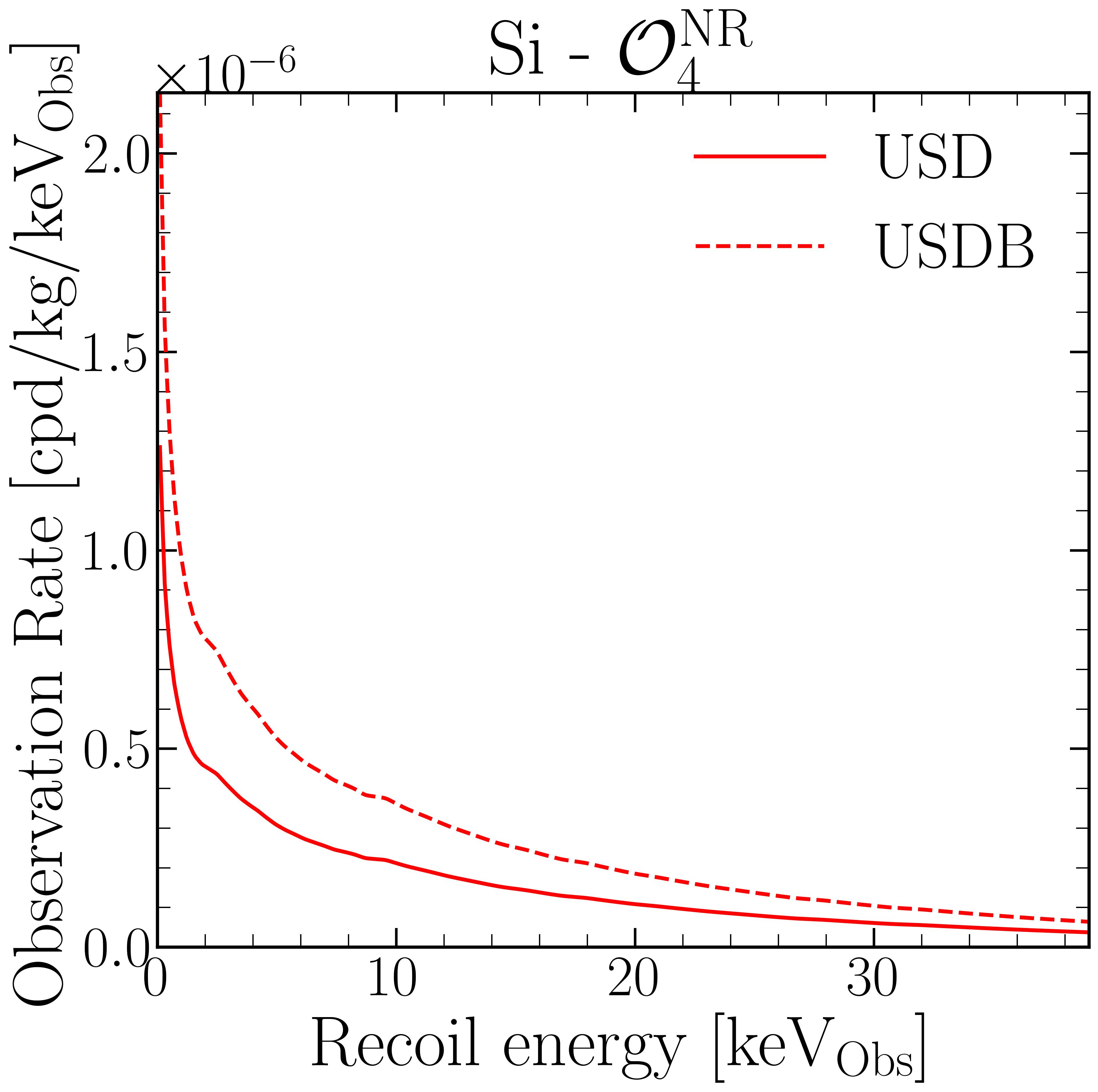}
\includegraphics[width=3.85cm,clip]{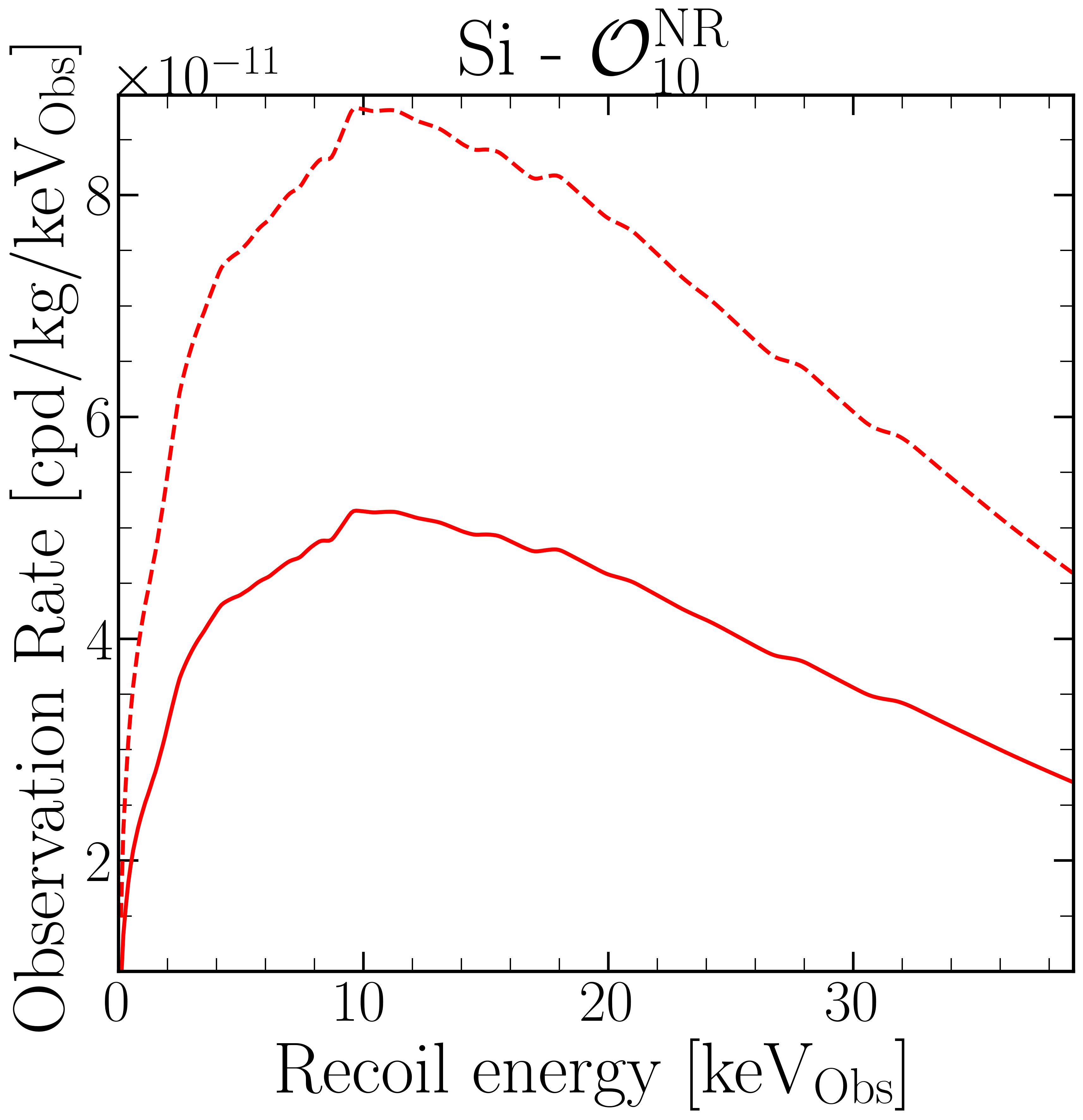}
\caption{Silicon observation rates for the $\mathcal{O}_4^{\rm NR}$ and $\mathcal{O}_{10}^{\rm NR}$ operators, for the shell model calculations USD (solid) and USDB (dashed), using $m_\chi=10$ GeV/$c^2$ and $\sigma_p=10^{-40}$ cm$^2$.}
\label{fig: si NR}      
\end{figure}

Figure~\ref{fig: si NR} showcases the $\mathcal{O}_4^{\rm NR}$ and $\mathcal{O}_{10}^{\rm NR}$ silicon observation rates, for the shell model calculations USD and USDB.  $\mathcal{O}_4^{\rm NR}$ constitutes the standard SD response, whilst $\mathcal{O}_{10}^{\rm NR}$ consists of the SD response $\Sigma''$ (see Table~\ref{table of NR and q operator nuclear}). In both cases the maximum nuclear difference in the rates is of order $\sim 70\%$, which is larger than that displayed by the neutron IFF silicon SD channels (of order $30-40\%$) \cite{AbdelKhaleq:2023ipt}. For silicon, the SD channels are significant in the analysis of nuclear uncertainties in scattering observables.

\begin{figure}[htpb]
\centering
\includegraphics[width=4.cm,clip]{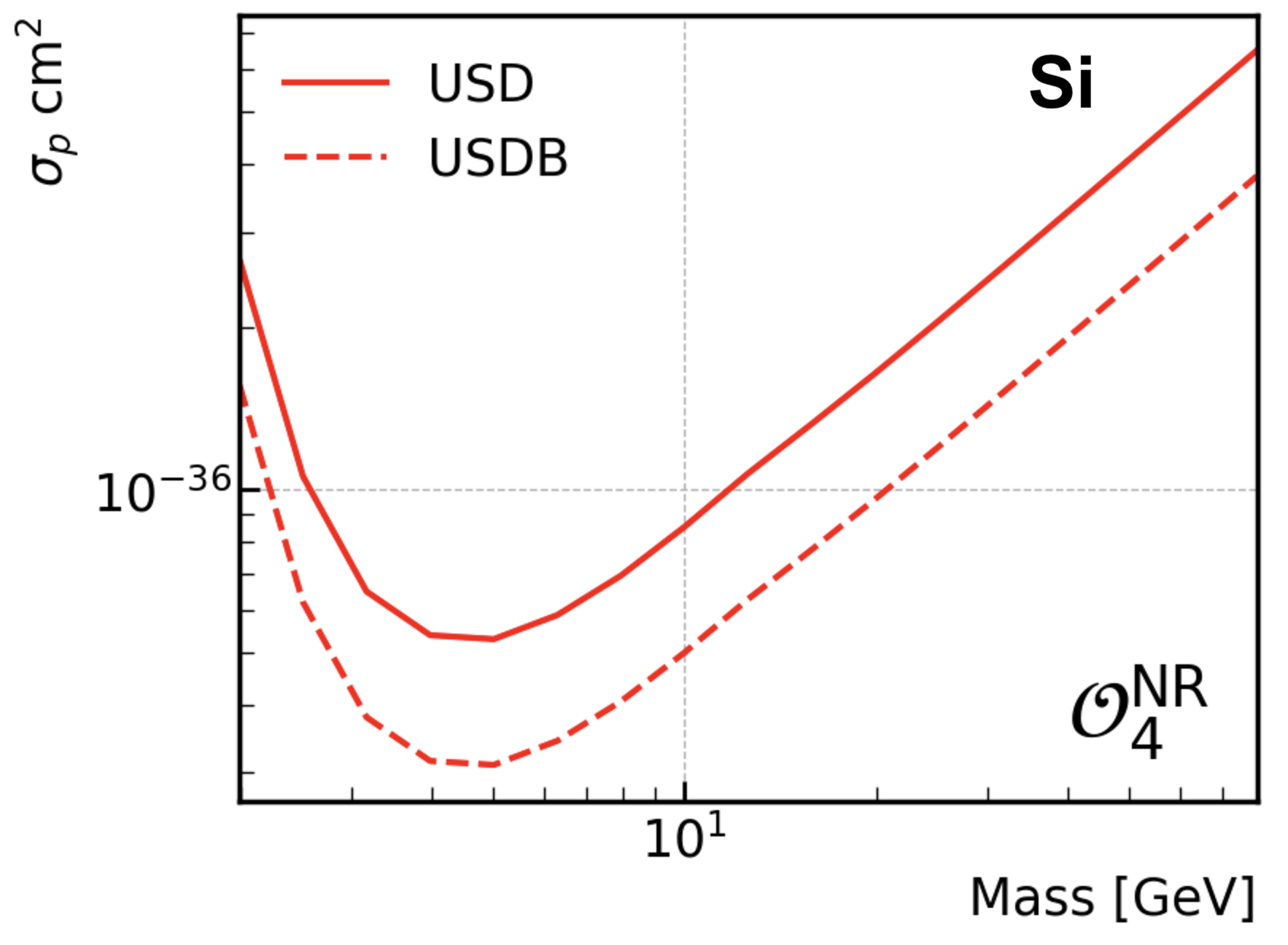}
\includegraphics[width=4.cm,clip]{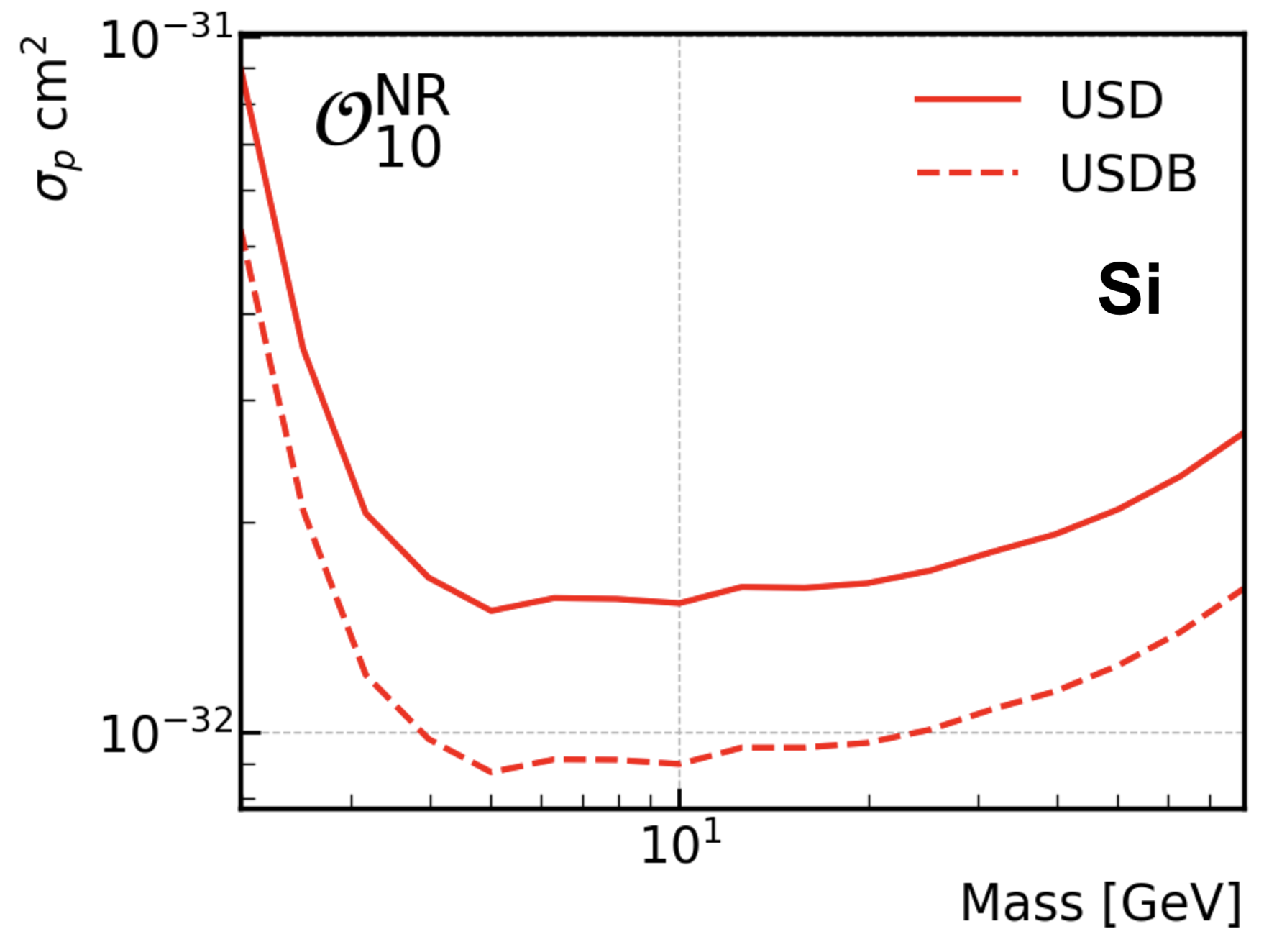}
\caption{SuperCDMS-like silicon $\mathcal{O}_4^{\rm NR}$ and $\mathcal{O}_{10}^{\rm NR}$ projection curves for two shell model calculations.}
\label{fig: si exclusion}   
\end{figure}

These nuclear differences at the scattering rate level are also reflected in the $\mathcal{O}_4^{\rm NR}$ and $\mathcal{O}_{10}^{\rm NR}$ exclusion limits presented in Fig.~\ref{fig: si exclusion}. Both $\mathcal{O}_4^{\rm NR}$ and $\mathcal{O}_{10}^{\rm NR}$ display a nuclear difference of order $75-80\%$ -- a similar value to that is seen in the scattering rates. In this case, the silicon experimental observables of interest are sensitive to nuclear shell model interaction choice.

\section{Conclusion}

Dark matter direct detection experiments aim to characterise the signals resulting from the scattering of WIMPs off nuclei, in order to better understand this elusive DM. Accurate modelling of the uncertainties present in the WIMP-nucleus elastic scattering rate is important for this characterisation and for experimental analysis. The scattering rate consists of several components which must be modelled and inputted, namely: the DM halo velocity distribution; the high energy physics coefficients describing the particle physics of interest; functions describing the experimental responses for the experiment in consideration; as well as nuclear form factors encoding the nuclear structure information of the target nuclei. \\

In this work we investigate the impact of nuclear modelling on WIMP-nucleus scattering rates and exclusion curves for silicon and germanium targets, quantifying the nuclear uncertainties present and the degree of propagation of these uncertainties from the nuclear form factors to experimental observables. This was done using nuclear shell model calculations employing various shell model interactions for each nuclear target, which were then compared and their differences quantified in the WIMP observables. Here, the DM halo velocity distribution is taken to be the SHM, and a NREFT framework was employed to describe the nuclear response functions, where velocity- and momentum-suppressed terms are considered in addition to the standard SI and SD responses. \\

The silicon and germanium scattering rates were plotted for a selection of NR operators $\mathcal{O}_i^{\rm NR}$ which showed non-negligible nuclear differences due to shell model interaction choice. These differences reflect those observed in the nuclear form factors, and can also be seen in the exclusion limits for these NR operators. As such, the nuclear uncertainties present at the form factor level are roughly retained at the scattering rate and exclusion limit levels, indicating a sensitivity of the WIMP scattering observables on nuclear modelling. This showcases the importance of improved and more accurate nuclear modelling for WIMP direct detection experimental analysis and interpretation.

\section*{Acknowledgements}
This research was supported by the Australian Government through the Australian Research Council Centre of Excellence for Dark Matter Particle Physics (CDM, CE200100008).

\bibliography{main.bib}

\begin{thebibliography}{19}

\bibitem{Fitzpatrick:2012ix}
A.L. Fitzpatrick, W.~Haxton, E.~Katz, N.~Lubbers, Y.~Xu, {The Effective Field Theory of Dark Matter Direct Detection}, JCAP \textbf{02}, 004 (2013), \texttt{1203.3542}. \doiwoc{10.1088/1475-7516/2013/02/004}

\bibitem{Anand:2013yka}
N.~Anand, A.L. Fitzpatrick, W.C. Haxton, {Weakly interacting massive particle-nucleus elastic scattering response}, Phys. Rev. C \textbf{89}, 065501 (2014), \texttt{1308.6288}. \doiwoc{10.1103/PhysRevC.89.065501}

\bibitem{Evans:2018bqy}
N.W. Evans, C.A.J. O'Hare, C.~McCabe, {Refinement of the standard halo model for dark matter searches in light of the Gaia Sausage}, Phys. Rev. D \textbf{99}, 023012 (2019), \texttt{1810.11468}. \doiwoc{10.1103/PhysRevD.99.023012}

\bibitem{OHare:2019qxc}
C.A.J. O'Hare, N.W. Evans, C.~McCabe, G.~Myeong, V.~Belokurov, {Velocity substructure from Gaia and direct searches for dark matter}, Phys. Rev. D \textbf{101}, 023006 (2020), \texttt{1909.04684}. \doiwoc{10.1103/PhysRevD.101.023006}

\bibitem{Brown:2014bhl}
B.A. Brown, W.D.M. Rae, {The Shell-Model Code NuShellX@MSU}, Nucl. Data Sheets \textbf{120}, 115 (2014). \doiwoc{10.1016/j.nds.2014.07.022}

\bibitem{Brown:1988vm}
B.A. Brown, B.H. Wildenthal, {Status of the nuclear shell model}, Ann. Rev. Nucl. Part. Sci. \textbf{38}, 29 (1988). \doiwoc{10.1146/annurev.ns.38.120188.000333}

\bibitem{Wildenthal:1984mf}
B.H. Wildenthal, {Empirical strengths of spin operators in nuclei}, Prog. Part. Nucl. Phys. \textbf{11}, 5 (1984). \doiwoc{10.1016/0146-6410(84)90011-5}

\bibitem{Brown:2006gx}
B.A. Brown, W.A. Richter, {New `USD' Hamiltonians for the sd shell}, Phys. Rev. C \textbf{74}, 034315 (2006). \doiwoc{10.1103/PhysRevC.74.034315}

\bibitem{AbdelKhaleq:2023ipt}
R.~Abdel~Khaleq, G.~Busoni, C.~Simenel, A.E. Stuchbery, {Impact of shell model interactions on nuclear responses to WIMP elastic scattering}, Phys. Rev. D \textbf{109}, 075036 (2024), \texttt{2311.15764}. \doiwoc{10.1103/PhysRevD.109.075036}

\bibitem{Honma:2009zz}
M.~Honma, T.~Otsuka, T.~Mizusaki, M.~Hjorth-Jensen, {New effective interaction for f5pg9-shell nuclei}, Phys. Rev. C \textbf{80}, 064323 (2009). \doiwoc{10.1103/PhysRevC.80.064323}

\bibitem{Mukhopadhyay:2017tca}
S.~Mukhopadhyay et~al., {Nuclear structure of Ge76 from inelastic neutron scattering measurements and shell model calculations}, Phys. Rev. C \textbf{95}, 014327 (2017). \doiwoc{10.1103/PhysRevC.95.014327}

\bibitem{Menendez:2008jp}
J.~Menendez, A.~Poves, E.~Caurier, F.~Nowacki, {Disassembling the Nuclear Matrix Elements of the Neutrinoless beta beta Decay}, Nucl. Phys. A \textbf{818}, 139 (2009), \texttt{0801.3760}. \doiwoc{10.1016/j.nuclphysa.2008.12.005}

\bibitem{AbdelKhaleq:2024hir}
R.~Abdel~Khaleq, J.L. Newstead, C.~Simenel, A.E. Stuchbery, {Detailed nuclear structure calculations for coherent elastic neutrino-nucleus scattering}, Phys. Rev. D \textbf{111}, 033003 (2025), \texttt{2405.20060}. \doiwoc{10.1103/PhysRevD.111.033003}

\bibitem{Aralis_2020}
T.~Aralis et~al. (SuperCDMS), {Constraints on dark photons and axionlike particles from the SuperCDMS Soudan experiment}, Phys. Rev. D \textbf{101}, 052008 (2020), [Erratum: Phys.Rev.D 103, 039901 (2021)], \texttt{1911.11905}. \doiwoc{10.1103/PhysRevD.101.052008}

\bibitem{supercdmscollaboration2023}
S.~Collaboration et~al., A strategy for low-mass dark matter searches with cryogenic detectors in the supercdms snolab facility (2023), \texttt{2203.08463}, \urlstyle{tt}\url{https://arxiv.org/abs/2203.08463}

\bibitem{Barker_2013}
D.~Barker, W.Z. Wei, D.M. Mei, C.~Zhang, {Ionization Efficiency Study for Low Energy Nuclear Recoils in Germanium}, Astropart. Phys. \textbf{48}, 8 (2013), \texttt{1304.6773}. \doiwoc{10.1016/j.astropartphys.2013.06.010}

\bibitem{si_yield}
M.F. Albakry et~al. ({SuperCDMS Collaboration}), {First Measurement of the Nuclear-Recoil Ionization Yield in Silicon at 100 eV}, {Phys. Rev. Lett.} \textbf{131}, 091801 (2023). \doiwoc{{10.1103/PhysRevLett.131.091801}}

\bibitem{Yellin:2002xd}
S.~Yellin, {Finding an upper limit in the presence of unknown background}, Phys. Rev. D \textbf{66}, 032005 (2002), \texttt{physics/0203002}. \doiwoc{10.1103/PhysRevD.66.032005}

\bibitem{ZurowskiGitHubSuperCDMS}
M.J. Zurowski, Signal generation and modelling (sgm), \url{https://github.com/mjzurowski/sgm/tree/dev/scdms_det} (2024)

\end{thebibliography}

\end{document}